\documentstyle[preprint,tighten,aps,epsfig]{revtex}
\begin{document}
\draft
%
%
%
%
\title{ Improved transport equations including correlations for
electron-phonon systems. Comparison with exact solutions in one dimension}
\author{J.~Fricke, V.~Meden, C.~W\"ohler, and K.~Sch\"onhammer}
\address{ Institut f\"{u}r Theoretische Physik der Universit\"{a}t
  G\"{o}ttingen, \\ 
Bunsenstra\ss{}e 9, 37073 G\"{o}ttingen, Germany}
\date{July 4, 1996}
\maketitle
\begin{abstract}
  We study transport equations for quantum many-particle systems in
  terms of correlations by applying the general formalism developed in
  an earlier paper to exactly soluble electron-phonon models. The
  one-dimensional models considered are the polaron model with a
  linear energy dispersion for the electrons and a finite number of
  electrons and the same model including a Fermi sea
  (Tomonaga-Luttinger model). The inclusion of two-particle
  correlations shows a significant and systematic improvement in
  comparison with the usual non-Markovian equations in Born
  approximation. For example, the improved equations take into account
  the renormalization of the propagation by the self-energies to
  second order in the coupling. 
\end{abstract}
\narrowtext
%
%
%
%
\section{Introduction}
\label{sec:Intro}
In a recent work \cite{PartI} one of the authors proposed a new method
for deriving transport equations in terms of correlations. In this
paper we test these transport equations and gain more insight into
their abilities to describe physical effects for some exactly soluble
electron-phonon models.

In their study of the Jaynes-Cummings model \cite{Zi} Zimmermann and
Wauer used the same decoupling procedure as ours one order higher than
the Born approximation in order to reduce unphysical results such as
fermionic occupation numbers outside the range $[\,0,1\,]$. We
extended this procedure systematically to all orders in the
correlations, thus obtaining an exact, infinite set of equations of
motion in terms of correlations. For this simple model it is easy to
solve these equations numerically to very high order in order to test
our method \cite{PhD}.

Here, we focus on one-dimensional versions of more realistic models,
i.e.\ the polaron model and the same model with a Fermi sea of
electrons (Tomonaga-Luttinger model). For the polaron model with a
linearized electronic energy dispersion exact results for
non-equilibrium expectation values were presented in Ref.\
\cite{MFWS}, as well as for the Tomonaga-Luttinger model with coupling
to phonons in Ref.\ \cite{MWFS}.  Similar models have recently been
discussed to study the relaxation of highly excited semiconductors and
insulators \cite{RZ,Ha,SKM,Da}. For these models non-Markovian
quantum-kinetic equations in Born approximation were numerically
solved assuming isotropic distributions in $\vec{k}$-space. It became
clear that damping factors for the electron propagation had to be
introduced in order to reduce the occurrence of unphysical results,
e.g.\ negative occupation numbers.  Renormalized free propagators
including the damping were obtained by the generalized Kadanoff-Baym
ansatz \cite{Li,Ha,KB}, which in turn was justified by Schoeller
\cite{Sch} as a partial resummation in a diagrammatic perturbation
theory. For the Tomonaga-Luttinger model including phonons it turned
out that the renormalization of the phonon frequencies plays a major
role for the relaxation process \cite{MWFS}.  In this work we study in
which way these important renormalizations of the free propagators
appear in our method, which contains only single-time quantities
without using advanced and retarded Green's functions. This is quite
interesting as the Hamiltonian dynamics is local in time and memory
effects only appear by truncating the infinite hierarchy of
quantities.

%
%
%
%
%

\section{The models}
\label{sec:model}

As in the first paper, we consider systems described by a general
statistical operator $\rho_0$ at the initial time $t_0$. The dynamics
is determined by a Hamiltonian $H$, and the expectation value of an
operator $A$ at time $t$ is given by $\langle A \rangle_t =
\mathrm{Tr} [\rho_0 A(t)]$, where $A(t)$ is the operator in the
Heisenberg representation with $A(t_0) = A$.

The one-dimensional polaron model is described by the following
Hamiltonian
\begin{equation}
  \label{eq:Polaron}
  H = \sum_k \epsilon_k \psi_{k}^{\dag}\psi_{k} + \sum_{q} \omega_q
  B_q^{\dag} 
  B_q + \sum_{q,k} g_q \left( \psi_{k+q}^{\dag}\psi_{k} B_q + B_q^{\dag}
  \psi_{k}^{\dag} \psi_{k+q} \right)
  \quad.
\end{equation}
The operator $\psi_k$ ($B_q$) annihilates an electron (phonon) in the
state with momentum $k$ ($q$) and energy $\epsilon_k$ ($\omega_q$).
The electron-phonon coupling strength is denoted by $g_q$.  In the
case of a linear energy dispersion $\epsilon_k = v_F k$ for the
electrons this model is exactly soluble for the initial
one-electron state $\rho_0 = \vert k_0 \rangle\langle k_0 \vert$
\cite{MFWS} or more generally for $\rho_0 = \vert k_1 ,\ldots ,k_N
\rangle\langle k_1 ,\ldots ,k_N \vert$ where $\vert k_1 ,\ldots ,k_N
\rangle$ is an anti-symmetrized $N$-electron state \cite{WS}. The time
evolution of the electron distribution function $n_k(t) = \langle
\psi_{k}^{\dag}\psi_{k}\rangle_t$ for the initial state $\rho_0 =
\vert k_0 \rangle\langle k_0 \vert$ can, e.g., be obtained by
numerically solving the exact differential equation derived in
Ref.~\cite{MFWS}:
\begin{equation}
  \label{eq:exactPol}
  \frac{d}{dt} n_k(t) = 2 \sum_{q>0} g_q^2 \frac{\sin\left[(\omega_q - v_F
    q)(t-t_0)\right]}{\omega_q - v_F q} \left[ n_{k+q}(t) - n_k(t) \right] 
  \quad.
\end{equation}
As in Ref.~\cite{MFWS} we assume that only phonons with $q>0$ exist in
accordance with the fact that each electron is moving ``right'' with
the velocity $v_F$ (chiral model).  Even for the initial state $\rho_0
= \vert k_0 ;\mathrm{FS} \rangle\langle k_0 ;\mathrm{FS} \vert$ ($k_0
> k_F$) with a Fermi sea $\vert {\mathrm{FS}} \rangle = \sum_{k \leq
  k_F} \psi_k^{\dag} \vert \mathrm{Vac} \rangle$ of electrons it turns
out that the exact time evolution of the electron and phonon
distribution functions can be numerically determined \cite{MWFS}. Here
we use the exact solutions for a comparison with the results of our
approximate transport equations.

The exact solutions are obtained for the case of an infinite range of
momenta $k= {2\pi n}/L$ with $ n= \ldots,-1,0,1,\ldots$ whereas the
approximate numerical calculations are for a system with a lower and
upper momentum cut-off ($ k_{<} \leq k \leq k_{>}$). These differences
are of little importance if the cut-offs are chosen to be far away
from the Fermi momentum resp.\ the initial state at $k_0$. In all
numerical calculations we use a constant electron-phonon coupling with
a momentum cut-off $q_c$, i.e.\ $g_q = g \, \Theta(q) \, \Theta
(q_c-q)$, and a constant phonon frequency $\omega_0\equiv \omega_q$ as
a model for longitudinal optical phonons.  We define the ``resonant''
phonon momentum $q_B$ by $\omega_0 = v_F q_B$ and choose $q_c \gg
q_B$. In the plots of the numerical results all momenta $k$ will be
given dimensionless as $k/q_B$ and all times $t$ as $\omega_0 t$.  The
system is described by the following dimensionless parameters: the
ratio of the one-electron level spacing and the phonon energy
$\nu=(v_F 2\pi/L)/\omega_{0} =(2\pi/L)/q_B$, the coupling constant
$\Gamma = g^2 /(\omega_{0} \cdot v_F 2\pi/L)$, and $q_c/q_B$.

%
%
%
%
%

\section{Quantum kinetic equations}
\label{sec:kinEq}

\begin{figure}[htbp]
  \begin{center}
    \leavevmode
    \epsfig{file=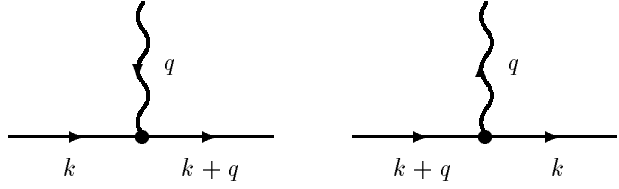}
  \end{center}
  \caption{The interaction vertices for the polaron model.}
  \label{fig:int_vertex}
\end{figure}

\begin{figure}[htbp]
  \begin{center}
    \leavevmode
    \epsfig{file=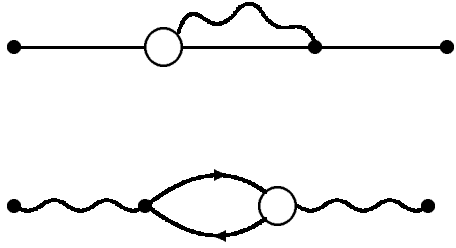}
  \end{center}
  \caption{The diagrams for the electron and the phonon distribution
    function.} 
  \label{fig:1part}
\end{figure}

\begin{figure}[htbp]
  \begin{center}
    \leavevmode
    \epsfig{file=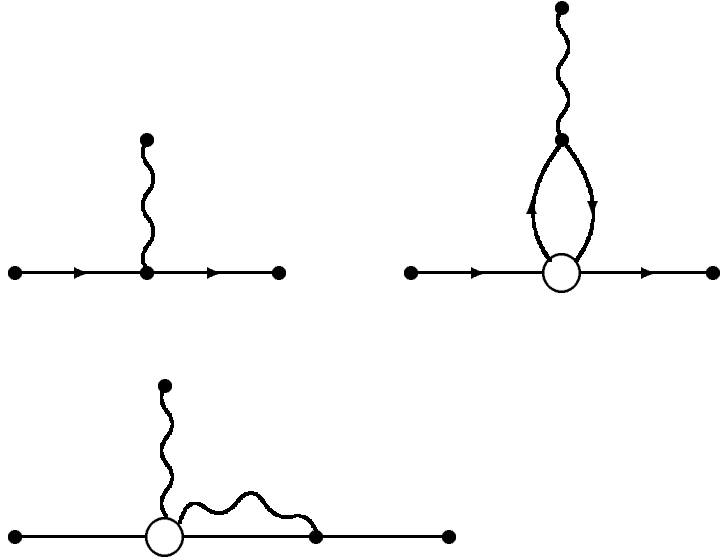}
  \end{center}
  \caption{The graphs for the time evolution of $h_{k,q}$ and $H_{k,q}$.}
  \label{fig:graph_hH}
\end{figure}

\begin{figure}[htbp]
  \begin{center}
    \leavevmode
    \epsfig{file=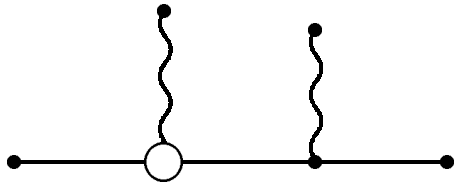}
  \end{center}
  \caption{The graph for the correlations $K_{k,q,q'}$, $k_{k,q,q'}$,
    $L_{k,q,q'}$, and $l_{k,q,q'}$.} 
  \label{fig:graph_kl}
\end{figure}

\begin{figure}[htbp]
  \begin{center}
    \leavevmode
    \epsfig{file=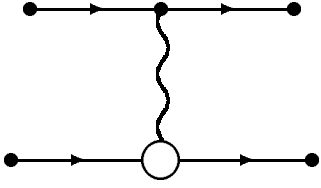}
  \end{center}
  \caption{The graph for the correlations $n_{k,q}$,
    $n_{k,q,q'}$, and $N_{k,q,q'}$.}
  \label{fig:graph_nN}
\end{figure}

We now present the equations of motion (EOM) for the correlations
derived using the general formalism described in Ref.~\cite{PartI}.
First, the transport equations one order beyond the Born approximation
are studied. The form of the equations is independent of $\rho_0$,
which only determines the initial values. The relevant correlations
are ($q,q'>0$):
\begin{eqnarray}
  n_k(t) & =& \langle \psi_{k}^{\dag}\psi_{k}\rangle_t  \\
  N_q(t) & =& \langle B_q^{\dag}B_q\rangle_t
  \quad,
\end{eqnarray}
\begin{eqnarray}
  h_{k,q}(t) & =& i\left( \langle
  B_q^{\dag}\psi_{k-q}^{\dag}\psi_{k}\rangle_t^c  
  - \langle B_q\psi_{k}^{\dag}\psi_{k-q}\rangle_t^c  \right) \\
  H_{k,q}(t) & =& \langle B_q^{\dag}\psi_{k-q}^{\dag}\psi_{k}\rangle_t^c 
  + \langle B_q\psi_{k}^{\dag}\psi_{k-q}\rangle_t^c  
  \quad,
\end{eqnarray}
\begin{eqnarray}
  n_{k,q}(t) &=& 2\langle
  \psi_{k-q}^{\dag}\psi_{k}^{\dag}\psi_{k-q}\psi_{k}\rangle_t^c  \\ 
  n_{k,q,q'}(t) &=& \langle
  \psi_{k-q-q'}^{\dag}\psi_{k}^{\dag}\psi_{k-q}\psi_{k-q'}\rangle_t^c
  +  
  \langle
  \psi_{k-q'}^{\dag}\psi_{k-q}^{\dag}\psi_{k}\psi_{k-q-q'}\rangle_t^c
  \\ 
  N_{k,q,q'}(t) &=& i
  \left( \langle
    \psi_{k-q-q'}^{\dag}\psi_{k}^{\dag}\psi_{k-q}\psi_{k-q'}\rangle_t^c
    - \langle
    \psi_{k-q'}^{\dag}\psi_{k-q}^{\dag}\psi_{k}\psi_{k-q-q'}\rangle_t^c
  \right) 
  \quad,
\end{eqnarray}
\begin{eqnarray}
  k_{k,q,q'}(t) &=& \langle
  B_q^{\dag}B_{q'}\psi_{k-q}^{\dag}\psi_{k-q'}\rangle_t^c  +  
  \langle B_{q'}^{\dag}B_q\psi_{k-q'}^{\dag}\psi_{k-q}\rangle_t^c  \\
  K_{k,q,q'}(t) &=& i\left( \langle
  B_q^{\dag}B_{q'}\psi_{k-q}^{\dag}\psi_{k-q'}\rangle_t^c  - 
  \langle B_{q'}^{\dag}B_q\psi_{k-q'}^{\dag}\psi_{k-q}\rangle_t^c  \right)
  \quad,
\end{eqnarray}
\begin{eqnarray}
  l_{k,q,q'}(t) &=& \langle
  B_q^{\dag}B_{q'}^{\dag}\psi_{k-q-q'}^{\dag}\psi_{k}\rangle_t^c  + 
  \langle B_{q'}B_q\psi_{k}^{\dag}\psi_{k-q-q'}\rangle_t^c  \\
  L_{k,q,q'}(t) &=& i\left(
  \langle B_q^{\dag}B_{q'}^{\dag}\psi_{k-q-q'}^{\dag}\psi_{k}\rangle_t^c 
  - \langle B_{q'}B_q\psi_{k}^{\dag}\psi_{k-q-q'}\rangle_t^c  \right)
  \quad.
\end{eqnarray}
The indices are chosen such that the correlations with three
indices are either symmetric or anti-symmetric in $q$ and $q'$. The
transport equations read
\begin{equation}
  \label{eq:EOM_nk}
  \frac{d}{dt} n_k(t) = \sum_{q>0} g_q \left( h_{k,q}(t)-h_{k+q,q}(t) \right)
  \quad,
\end{equation}
\begin{equation}
  \label{eq:EOM_Nq}
  \frac{d}{dt} N_q(t) = - g_q \sum_k h_{k,q}(t)
  \quad,
\end{equation}
\begin{eqnarray}
  \label{eq:EOM_hkq}
  \frac{d}{dt} h_{k,q}(t) &=& (\epsilon_k -\epsilon_{k-q}-\omega_q)H_{k,q}(t)
  \\ \nonumber 
  && +2 g_q \left[ N_q(t)\left( n_{k-q}(t)-n_k(t)\right) -
  n_k(t)\left(1 -n_{k-q}(t)\right)\right]
  \\ \nonumber  &&
  - g_q [ n_{k,q}(t) + \sum_{q'>0} ( n_{k,q,q'}(t) +
  n_{k+q',q,q'}(t) ) ]
  \\ \nonumber  &&
  +\sum_{q'>0} g_{q'} \left( k_{k,q,q'}(t) - k_{k+q',q,q'}(t)  -
  l_{k,q,q'}(t) + l_{k+q',q,q'}(t) \right)
  \quad,
\end{eqnarray}
\begin{eqnarray}
  \label{eq:EOM_Hkq}
  \frac{d}{dt} H_{k,q}(t) &=& -(\epsilon_k -\epsilon_{k-q}-\omega_q)h_{k,q}(t)
  \\ \nonumber  &&
  - g_q \sum_{q'>0} \left( N_{k,q,q'}(t) - N_{k+q',q,q'}(t) \right)
  \\ \nonumber  && 
  -\sum_{q'>0} g_{q'} \left( K_{k,q,q'}(t) - K_{k+q',q,q'}(t)  -
  L_{k,q,q'}(t) + L_{k+q',q,q'}(t) \right)
  \quad,
\end{eqnarray}
\begin{equation}
  \frac{d}{dt} n_{k,q}(t) = 2 g_q \left( n_{k-q}(t) -n_k(t)\right) h_{k,q}(t)
  \quad,
\end{equation}
\begin{eqnarray}
  \frac{d}{dt} n_{k,q,q'}(t) &=& (\epsilon_{k-q-q'}+\epsilon_k -\epsilon_{k-q}
  -\epsilon_{k-q'}) N_{k,q,q'}(t)
  \\ \nonumber  &&
  +
  \left\{ g_q
    \left[ \left(n_{k-q-q'}(t)-n_{k-q'}(t)\right)
      h_{k,q}(t) + \left(n_{k-q}(t)-n_{k}(t)\right) h_{k-q',q}(t)
    \right] 
  \right.
  \\ \nonumber  &&
  \left.
    - (q\leftrightarrow q')
  \right\}
  \quad,
\end{eqnarray}
\begin{eqnarray}
  \frac{d}{dt} N_{k,q,q'}(t) &=& -(\epsilon_{k-q-q'}+\epsilon_k -\epsilon_{k-q}
  -\epsilon_{k-q'}) n_{k,q,q'}(t) 
  \\ \nonumber  &&
  +
  \left\{ g_q
    \left[ \left(n_{k-q-q'}(t)-n_{k-q'}(t)\right)
      H_{k,q}(t) - \left(n_{k-q}(t)-n_{k}(t)\right) H_{k-q',q}(t)
    \right]
  \right.
  \\ \nonumber  &&
  \left.
    - (q\leftrightarrow q')
  \right\}
  \quad,
\end{eqnarray}
\begin{eqnarray}
  \frac{d}{dt} k_{k,q,q'}(t) &=& (\epsilon_{k-q}+\omega_{q}-\epsilon_{k-q'}
  -\omega_{q'}) K_{k,q,q'}(t)
  \\ \nonumber  &&
  +
  \left\{ g_q
    \left[ \left( N_q(t)+n_{k-q'}(t)\right) h_{k-q,q'}(t) 
      -\left( 1+N_q(t)-n_{k-q}(t)\right) h_{k,q'}(t) 
    \right]
  \right.
  \\ \nonumber  &&
  \left.
    + (q\leftrightarrow q')
  \right\} 
  \quad,
\end{eqnarray}
\begin{eqnarray}
  \frac{d}{dt} K_{k,q,q'}(t) &=& -(\epsilon_{k-q}+\omega_{q}-\epsilon_{k-q'}
  -\omega_{q'}) k_{k,q,q'}(t)
  \\ \nonumber  &&
  + 
  \left\{ g_q
    \left[ \left( N_q(t)+n_{k-q'}(t)\right) H_{k-q,q'}(t) 
      -\left( 1+N_q(t)-n_{k-q}(t)\right) H_{k,q'}(t) 
    \right]
  \right. 
  \\ \nonumber  &&
  \left.
    - (q\leftrightarrow q')
  \right\}
  \quad,
\end{eqnarray}
\begin{eqnarray}
  \frac{d}{dt} l_{k,q,q'}(t) &=& (\omega_q+\omega_{q'}+\epsilon_{k-q-q'}
  -\epsilon_k) L_{k,q,q'}(t)
  \\ \nonumber  &&
  +
  \left\{ g_q
    \left[ -\left( N_q(t)+n_{k}(t)\right) h_{k-q,q'}(t) 
      +\left( 1+N_q(t)-n_{k-q-q'}(t)\right) h_{k,q'}(t) 
    \right]
  \right.
  \\ \nonumber  &&
  \left.
    + (q\leftrightarrow q')
  \right\}
  \quad,
\end{eqnarray}
\begin{eqnarray}
  \label{eq:EOM_L}
  \frac{d}{dt} L_{k,q,q'}(t) &=& -(\omega_q+\omega_{q'}+\epsilon_{k-q-q'}
  -\epsilon_k) l_{k,q,q'}(t)
  \\ \nonumber  &&
  +
  \left\{ g_q
    \left[ \left( N_q(t)+n_{k}(t)\right) H_{k-q,q'}(t) 
      -\left( 1+N_q(t)-n_{k-q-q'}(t)\right) H_{k,q'}(t) 
    \right]
  \right. 
  \\ \nonumber  &&
  \left.
    + (q\leftrightarrow q')
  \right\}
  \quad.
\end{eqnarray}
In the graphical representation of the EOM \cite{PartI} we often use
fermionic (solid) and bosonic (curly) lines without arrows in order to
reduce the number of diagrams. Arrows and momentum labels are supposed
to be added according to the conservation of the number of electrons
and of the momentum. The interaction vertices are given in
Fig.~\ref{fig:int_vertex}.  The graphs for the distribution functions
$n_k$ and $N_q$ are those shown in Fig.~\ref{fig:1part}, the diagrams
for the ``phonon assisted polarizations'' $h_{k,q}$ and $H_{k,q}$ are
shown in Fig.~\ref{fig:graph_hH}, the graph for $k_{k,q,q'}$,
$K_{k,q,q'}$, $l_{k,q,q'}$ and $L_{k,q,q'}$ is given in
Fig.~\ref{fig:graph_kl} and the graph for $n_{k,q}$, $n_{k,q,q'}$ and
$N_{k,q,q'}$ is the one shown in Fig.~\ref{fig:graph_nN}.

By restriction to the correlations $n_k$, $N_q$, $h_{k,q}$ and
$H_{k,q}$ the transport equations in Born approximation are obtained.
For the case of vanishing initial correlations they read after
integrating the differential equations for $h_{k,q}$ and $H_{k,q}$:
\begin{eqnarray}
  \label{eq:Born}
  \frac{d}{dt} n_k(t) &=& 2 \sum_{q>0} g_q^2 \int_{t_0}^{t} dt'
  \left\{
    \cos\left((\epsilon_k- \epsilon_{k-q} -\omega_q)(t-t') \right) 
  \right.
  \\ \nonumber  && \times
  \left[ N_q  \left( n_{k-q}-n_k \right) 
    - n_k\left( 1-n_{k-q} \right) 
  \right]_{t'}
  \\ \nonumber  &&
  \left.
    - \cos\left((\epsilon_{k+q}- \epsilon_{k} -\omega_q)(t-t') \right) 
    \left[ N_q  \left( n_{k}-n_{k+q} \right) 
      - n_{k+q}\left( 1-n_{k} \right) 
    \right]_{t'}
  \right\}
  \quad,
\end{eqnarray}
\begin{eqnarray}
  \label{eq:Born_}
  \frac{d}{dt} N_q(t) &=& -2 g_q^2 \sum_k \int_{t_0}^{t} dt'
  \cos\left((\epsilon_k- \epsilon_{k-q} -\omega_q)(t-t') \right) 
  \\ \nonumber  && \times
  \left[ N_q  \left( n_{k-q}-n_k \right) 
    - n_k\left( 1-n_{k-q} \right) 
  \right]_{t'}
  \quad.
\end{eqnarray}

%
%
%
%
%

\subsection{The polaron model}
\label{sec:pol}

We first discuss the transport equations for the one-electron polaron
model with the initial state $\rho_0=| k_0\rangle \langle k_0 |$. A
renormalization of the free propagators, i.e.\ the cosine factors in
Eqs.~(\ref{eq:Born}) and (\ref{eq:Born_}), is of importance for the
Born approximation to yield acceptable results for the first phonon
replica at larger times \cite{MWFS,Ha}. In the one-electron case there
is no significant renormalization of the phonon propagators and the
most important renormalization is the inclusion of a damping factor
$\exp [-2\gamma (t-t') ]$ for the electronic propagations. For the
determination of $\gamma$ from the second order electron self-energy
in equilibrium we refer to Ref.~\cite{MWFS}.

\begin{figure}[htbp]
  \begin{center}
    \leavevmode
    \epsfig{file=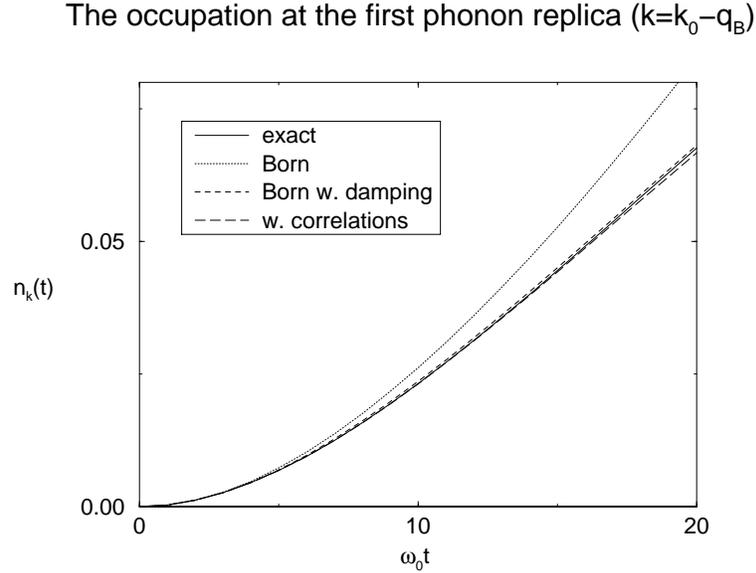,height=10cm,angle=270}
  \end{center}
  \caption{Time evolution of the electron occupation at the first
    phonon replica ($k=k_0-q_B$) for the one-electron polaron system
    with $\nu = 1/{16}$, $\Gamma = 0.005$, and $q_c/q_B= 5$. The exact
    solution is compared with the results from the kinetic equations
    in Born approximation with and without damping of the electron
    propagation and the improved equations including two-particle
    correlations.}
  \label{fig:pol_1rep}
\end{figure}

We solved numerically the equations including the two-particle
correlations as described in Eqs.~(\ref{eq:EOM_nk})-(\ref{eq:EOM_L}).
The initial states discussed so far do not have initial correlations.
Fig.~\ref{fig:pol_1rep} shows the time evolution of $n_k$ for the
first phonon replica at $k=k_0 - q_B$ with $q_B = \omega_0 /v_F $. The
importance of the damping of the electronic propagation in the Born
approximation can be seen clearly. The approximation including the
two-particle correlations seems to include the damping in the dynamics
as it follows the exact evolution for a long time.

\begin{figure}[htbp]
  \begin{center}
    \leavevmode
    \epsfig{file=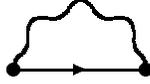}
  \end{center}
  \caption{Second order self-energy for the electron propagator.}
  \label{fig:el_renorm}
\end{figure}

We now explain how the dynamical equations incorporate the damping of
the electron propagation. In the Keldysh formalism \cite{Ra} the
renormalization of the free propagators is usually justified by the
Kadanoff-Baym ansatz \cite{Li,KB,Sch}, which substitutes the free
propagators in Eqs.~(\ref{eq:Born}) and (\ref{eq:Born_}) by the exact
retarded/advanced Green's functions. As these Green's functions are
not known they are, in a first approximation, taken in lowest order
perturbation theory for the equilibrium. In our case, the electron
propagators are renormalized by the self-energy in second order Born
approximation shown in Fig.~\ref{fig:el_renorm}. Diagrammatically
these self-energy insertions arise from the third diagram in
Fig.~\ref{fig:graph_hH}. Therefore, in order to explain the
renormalization we can neglect for a moment the two-electron
correlations $n_{k,q}$, $n_{k,q,q'}$ and $N_{k,q,q'}$. Then the
differential equations for the phonon assisted density matrices read
\begin{eqnarray}
  \label{eq:damp_start}
  \lefteqn{\left(\frac{d}{dt} +i(\epsilon_k -\epsilon_{k-q}-\omega_q)\right)
    (h_{k,q} +i H_{k,q})(t)}
  \\ \nonumber{}
  &=& 
  2 g_q \left[ N_q(t)\left( n_{k-q}(t)-n_k(t)\right) -
  n_k(t)\left(1 -n_{k-q}(t)\right)\right]
  \\ \nonumber  &&
  +\sum_{q'>0} g_{q'} \left( (k_{k,q,q'} -i K_{k,q,q'})(t) -
  (k_{k+q',q,q'} -i K_{k+q',q,q'})(t) \right.
  \\ \nonumber{}  && 
  \left.
    -(l_{k,q,q'} -i L_{k,q,q'})(t) +(l_{k+q',q,q'} -i L_{k+q',q,q'})(t)
  \right) 
  \quad.
\end{eqnarray}
The terms in the time derivative of $(k_{\dots} -i K_{\dots})$ and 
$(l_{\dots} -i L_{\dots})$ that yield the renormalization of the
electron propagation for $(h_{k,q} +i H_{k,q})$  are the
``coherent terms'', i.e.\ those proportional to $h_{k,q}$ and $H_{k,q}$:
\begin{eqnarray}
  \lefteqn{\left(\frac{d}{dt}-i (\epsilon_{k-q}+\omega_{q}-\epsilon_{k-q'}
    -\omega_{q'})\right) (k_{k,q,q'} -i K_{k,q,q'})(t)}\\ 
  \nonumber{} &=& 
  - g_{q'}
  \left( 1+N_{q'}(t)-n_{k-q'}(t)\right) (h_{k,q} +i H_{k,q})(t)  
    + \ldots
\end{eqnarray}
\begin{eqnarray}
  \lefteqn{\left(\frac{d}{dt}-i (\epsilon_{k+q'-q}+\omega_{q}-\epsilon_{k}
    -\omega_{q'})\right) (k_{k+q',q,q'} -i K_{k+q',q,q'})(t)}\\ 
  \nonumber{} &=& 
  g_{q'}
  \left( N_{q'}(t)+n_{k+q'-q}(t)\right) (h_{k,q} +i H_{k,q})(t)  
    + \ldots
\end{eqnarray}
and the analogous expressions for $(l_{\dots} -i L_{\dots})$. If
there are no initial correlations we obtain the following
integro-differential equation for $(h_{k,q} +i H_{k,q})$
neglecting the ``incoherent terms'':
\begin{eqnarray}
  \label{eq:phonass_renorm}
  \lefteqn{\left(\frac{d}{dt} +i(\epsilon_k -\epsilon_{k-q}-\omega_q)\right)
    (h_{k,q} +i H_{k,q})(t)}\\ 
  \nonumber{} &=&
  2 g_q \left[ N_q(t)\left( n_{k-q}(t)-n_k(t)\right) -
  n_k(t)\left(1 -n_{k-q}(t)\right)\right] \\ 
  \nonumber 
  &-& \int_{t_0}^t dt'\, \Pi_{k,q}(t,t')(h_{k,q} +i H_{k,q})(t')
  \quad,  
\end{eqnarray}
where 
\begin{eqnarray}
  \label{eq:self-energy}
  \Pi_{k,q}(t,t') &=& \Theta(t-t') \sum_{q'>0} g_{q'}^2
  \left[
    \left( 1+N_{q'}-n_{k-q'}
    \right)(t')
    e^{i(\epsilon_{k-q}+\omega_{q}-\epsilon_{k-q'} 
      -\omega_{q'})(t-t')}
  \right.
  \\ \nonumber{} &+&
  \left(N_{q'}+n_{k+q'}
  \right)(t')
  e^{i(\omega_q+\omega_{q'}+\epsilon_{k-q}
    -\epsilon_{k+q'})(t-t')}
  \\ \nonumber{} &+&
  \left(1 + N_{q'}-n_{k-q-q'}
  \right)(t')
  e^{i(\omega_q+\omega_{q'}+\epsilon_{k-q-q'}
    -\epsilon_k)(t-t')}
  \\ \nonumber{} &+&
  \left.
    \left( N_{q'}+n_{k+q'-q}
    \right)(t')
    e^{i(\epsilon_{k+q'-q}+\omega_{q}-\epsilon_{k}
      -\omega_{q'})(t-t')}
  \right]
  \quad.
\end{eqnarray}
In this context, the time-dependence of the distribution functions
$n_k$ and $N_q$ is of little importance, as can be checked
numerically. If we therefore replace them by their initial values
$n_k(t_0)$ and $N_q(t_0)$, the ``self-energy'' $\Pi_{k,q}(t,t')$ depends
only on the relative time $t-t'$. Then the differential equation
(\ref{eq:phonass_renorm}) is easily solved:
\begin{equation}
  \label{eq:phonass_sol}
  (h_{k,q}+i H_{k,q})(t) =  2 g_q \int_{t_0}^t dt'\, D_{k,q} (t-t') 
  \left[ N_q(t')\left( n_{k-q}(t')-n_k(t')\right) -
  n_k(t')\left(1 -n_{k-q}(t')\right)\right]
\end{equation}
where $D_{k,q}$ is the retarded solution to the Dyson equation
\begin{equation}
  \label{eq:Dyson}
  D_{k,q}(t-t') = D^{(0)}_{k,q}(t-t') -  \int_{t'}^t dt_1\,
  \int_{t'}^{t_1} dt_2\,  D^{(0)}_{k,q}(t-t_1)\Pi_{k,q}(t_1-t_2)
  D_{k,q}(t_2-t') 
\end{equation}
with the ``free propagation'' $D^{(0)}_{k,q}(t-t') = \Theta(t-t')
e^{-i(\epsilon_k-\epsilon_{k-q} -\omega_q)(t-t')}$.  Except for the
fact that $D_{k,q}$ is the ``propagator'' for an electron-hole pair
plus a phonon, this is an ordinary Dyson equation, which can be easily
solved by Fourier transformation
\begin{equation}
  \label{eq:Dyson_sol}
  -i \widetilde{D}_{k,q}(\omega) = 
  \left( \omega - (\epsilon_k-\epsilon_{k-q} -\omega_q) +i
    \widetilde{\Pi}_{k,q}(\omega) +i0 
  \right)^{-1}
\end{equation}
and leads to a damping for both electronic propagations.

\begin{figure}[htbp]
  \begin{center}
    \leavevmode
    \epsfig{file=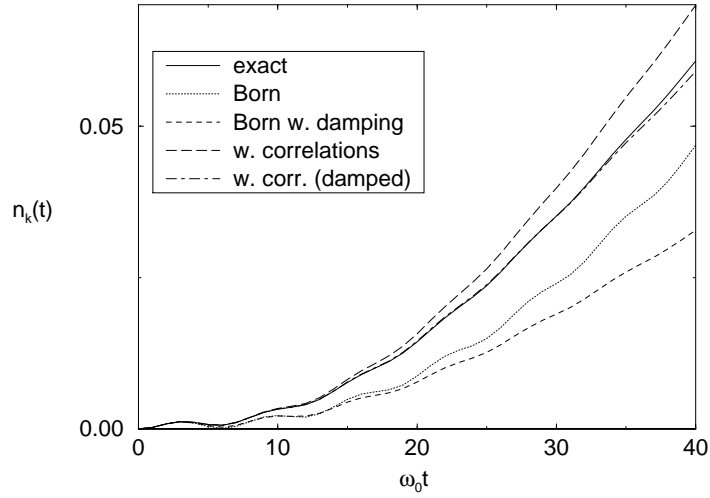,height=10cm,angle=270}
  \end{center}
  \caption{Time evolution of the electron occupation at the second
    phonon replica ($k=k_0-2 q_B$) for the system of
    Fig.~\ref{fig:pol_1rep}. In addition to Fig.~\ref{fig:pol_1rep}
    the improved equations with damping are considered.
    }
  \label{fig:pol_2rep}
\end{figure}

As can be seen from Fig.~\ref{fig:pol_2rep} the time evolution of the
occupation at the second phonon replica at $k=k_0 -2 q_B$ is much
better described by the approximate equations including correlations
(long dashed line). This is easily explained by the fact that contrary
to the Born approximation these equations contain coherent two-phonon
processes by the inclusion of expectation values $\langle B^{(\dag)}
B^{(\dag)} \psi^{\dag} \psi \rangle_t^c$. By an analogous argument to
the one used to explain a damping factor in the Born approximation it
can be justified to damp the time evolution of $n_{k,q}$,
$n_{k,q,q'}$, $N_{k,q,q'}$, $k_{k,q,q'}$, $K_{k,q,q'}$, $l_{k,q,q'}$,
and $L_{k,q,q'}$, i.e.\ replacing $\frac{d}{dt}$ by $\left(
\frac{d}{dt} +2\gamma\right)$ in the corresponding differential
equations. Why the correlations with four electron operators $\langle
\psi^{\dag} \psi^{\dag} \psi\psi \rangle_t^c$ are damped with
$2\gamma$ instead of $4\gamma$ as one might expect will be discussed
later when we extend the equations by ``half an order''. The inclusion
of this damping for the two-particle correlations yields an
improvement of the solution for longer times in the same way as the
damping of the Born approximation does one order higher (dot-dashed
line in Fig.~\ref{fig:pol_2rep}).

\begin{figure}[htbp]
  \begin{center}
    \leavevmode
    \epsfig{file=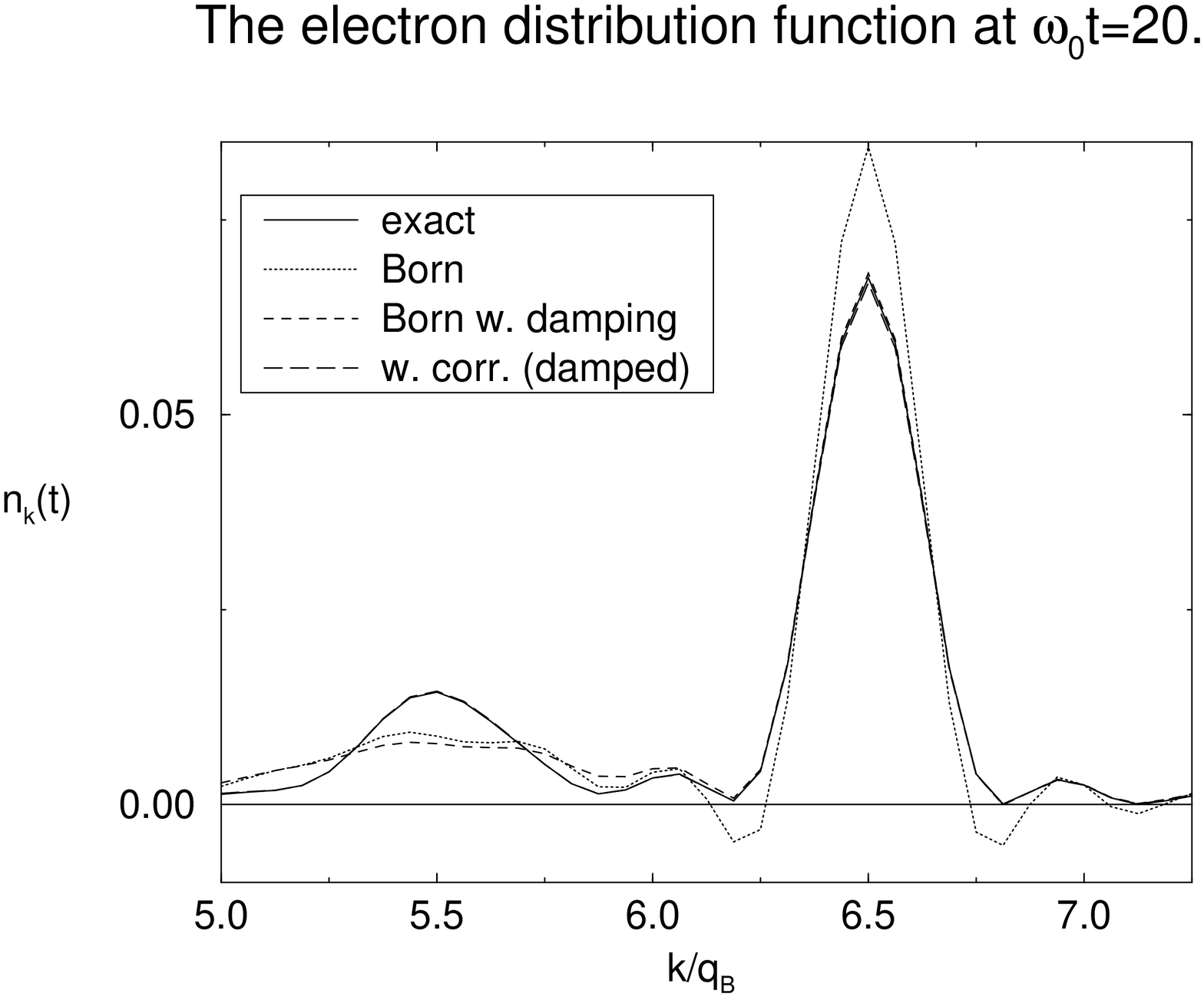,height=10cm,angle=270}
  \end{center}
  \caption{The electron distribution function at $\omega_0 t=20$ for
    the system of Fig.~\ref{fig:pol_1rep}. Shown are the first and
    second phonon replica. The electron was initially at $k_0= 7.5 q_B$.} 
  \label{fig:pol_distr10}
\end{figure}

\begin{figure}[htbp]
  \begin{center}
    \leavevmode
    \epsfig{file=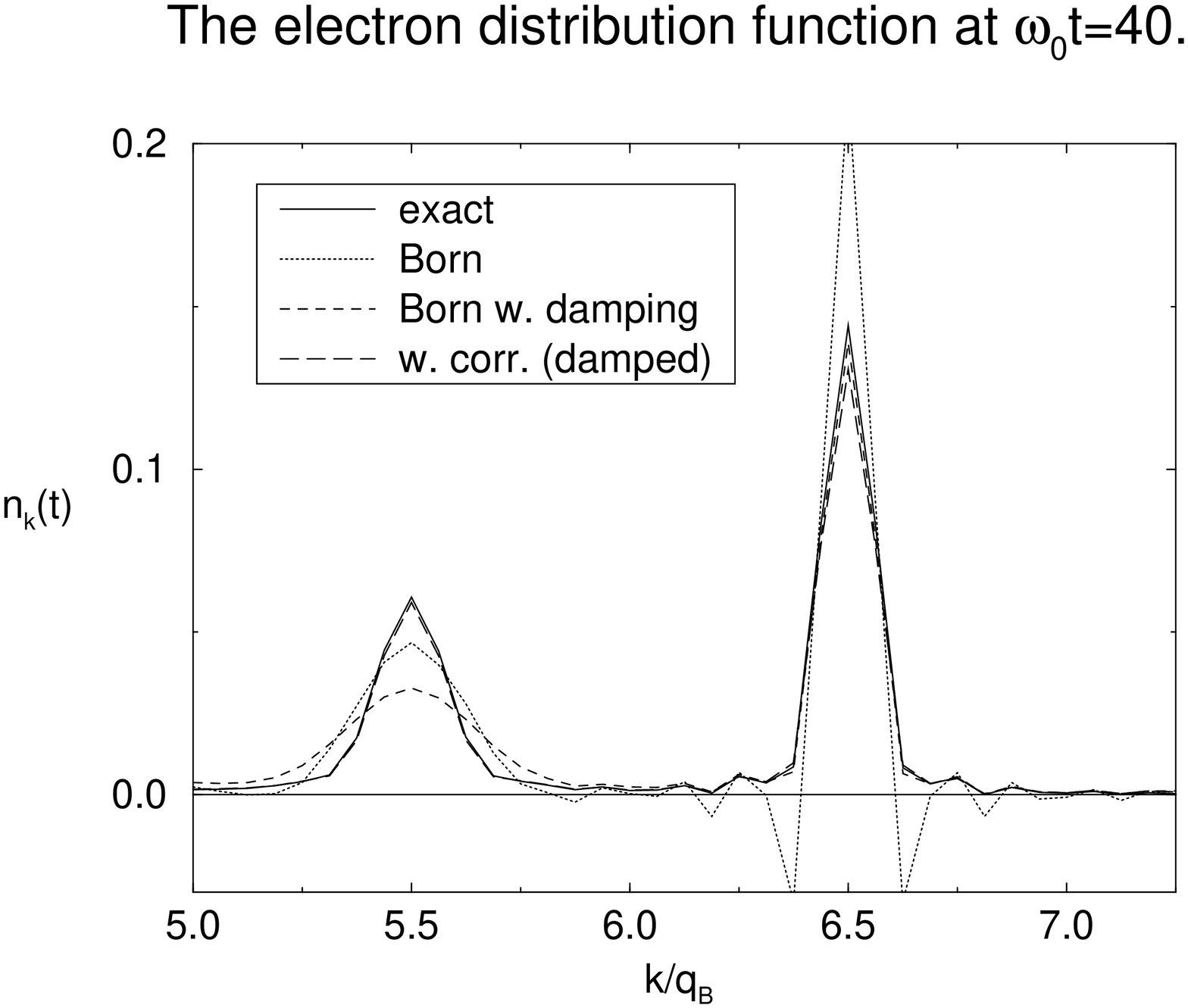,height=10cm,angle=270}
  \end{center}
  \caption{The same as in Fig.~\ref{fig:pol_distr10}, but for the time
     $\omega_0 t=40$.}
  \label{fig:pol_distr20}
\end{figure}

Figs.~\ref{fig:pol_distr10} and \ref{fig:pol_distr20} show the
electron distribution function at two different times. In the range of
the first phonon replica, the distribution function is well described
by the Born approximation with a damping factor for the free electron
propagation (dashed lines) and by the kinetic equations including
two-particle correlations (long dashed lines), whereas the Born
approximation without damping (dotted lines) yields unphysical results
like negative occupation numbers. For the second phonon replica only
the kinetic equations including two-particle correlations are in
excellent agreement with the exact solution (solid lines).

%
%
%
%
%

\subsection{The Tomonaga-Luttinger model}
\label{sec:TL}

We now turn to the model with a Fermi sea (Tomonaga-Luttinger model).
The initial state $\rho_0 = \vert k_0 ;\mathrm{FS} \rangle\langle k_0
;\mathrm{FS} \vert$ again contains no correlations. The most important
effect of the Fermi sea (apart form the Pauli blocking) is the
renormalization of the phonon propagator. Due to the linear energy
dispersion of the electrons, which also enables the exact solution by
bosonization \cite{MWFS}, the phonon mode splits into two modes in the
equilibrium case and the second order self-energy (electron-hole
bubble) already gives the exact result for the equilibrium phonon
Green's function. In Ref.~\cite{MWFS} the importance of the phonon
renormalization for the kinetic equations in Born approximation was
shown. In our dynamical equations this renormalization is incorporated
by the second diagram in Fig.~\ref{fig:graph_hH}. By extending the
discussion following Eq.~\ref{eq:damp_start} to the case of the
renormalization of both the electron and phonon propagation, i.e.\ by
inclusion of the electron correlations $n_{k,q}$, $n_{k,q,q'}$, and
$N_{k,q,q'}$, we obtain the ``self-energy''
\begin{equation}
  \label{eq:pi_wrong}
  -i \widetilde{\Pi}_{k,q}(\omega) = \frac{g_q^2 \sum_{k'}
    (n_{k'-q}-n_{k'})(t_0)} {\omega +i0} -i2\gamma
\end{equation}
for the ``propagator'' $\widetilde{D}_{k,q}(\omega)$ in
Eq.~(\ref{eq:Dyson_sol}), where we approximated the electron
self-energies by a constant damping factor.  But this termination of
the ``continued fraction expansion'' for the propagator
$\widetilde{D}_{k,q}(\omega)$ is not the best one. Better results are
obtained if the damping of the propagation of the scattered electron
in $n_{k,q},n_{k,q,q'}$, and $N_{k,q,q'}$ is included:
\begin{equation}
  \label{eq:pi_right}
  -i \widetilde{\Pi}_{k,q}(\omega) = \frac{g_q^2 \sum_{k'}
    (n_{k'-q}-n_{k'})(t_0)} {\omega +i2\gamma} -i2\gamma
  \quad.
\end{equation}
Now $\sum_{k'} (n_{k'-q}-n_{k'})(t_0) = \frac{L}{2\pi} q$ for $0<q\leq
q_c$, so that the propagator $D_{k,q}$ is given as described in
Ref.~\cite{MWFS} by
\begin{equation}
  \label{eq:phon_modes}
  D_{k,q}(t-t') = \Theta(t-t')
  \left\{ \frac{\lambda_{+}}{\lambda_{+}-\lambda_{-}}
    e^{(-i\lambda_{+} -2\gamma)(t-t')} -
    \frac{\lambda_{-}}{\lambda_{+}-\lambda_{-}} 
    e^{(-i\lambda_{-} -2\gamma)(t-t')}
  \right\}
\end{equation}
where
\begin{equation}
  \label{eq:lamda}
  \lambda_{\pm} = \frac{1}{2} 
  \left\{ (v_F q -\omega_q) \pm \sqrt{ (v_F q -\omega_q)^2 + 4  g_q^2
      \frac{L}{2\pi} q}
  \right\}
  \quad.
\end{equation}

The splitting of the phonon frequency into two modes as well as the
above mentioned cancellation of certain corrections for the
electron-hole bubble are special effects of the linear electronic
energy dispersion. The terms proportional to $k_{k,q,q'}$,
$K_{k,q,q'}$, $l_{k,q,q'}$, and $L_{k,q,q'}$ in the time evolution of
the phonon assisted density matrices $h_{k,q}$ and $H_{k,q}$
(Eqs.~(\ref{eq:EOM_hkq})-(\ref{eq:EOM_Hkq})), which led to a damping
for $h_{k,q}$ and $H_{k,q}$, do not contribute to the time evolution
of $N_q$ (Eq.~(\ref{eq:EOM_Nq})) due to the summation of $h_{k,q}$
over $k$. (There are contributions from electron states near the
momentum cut-offs which do not cancel exactly. But in the spirit
discussed above they may be neglected.)  This cancellation occurs only
due to the linear energy dispersion which makes the non-interacting
energies $\epsilon_{k} -\epsilon_{k-q} -\omega_q = v_F q -\omega_q$ of
$h_{k,q}$ and $H_{k,q}$ independent of $k$.  Therefore, the damping of
$h_{k,q}$ and $H_{k,q}$ is only ``existent'' for their appearance in
the differential equation (\ref{eq:EOM_nk}) for $n_k$, but not in that
for $N_q$. Although for the parameters chosen this remark is of little
importance, it should have been taken into account for the equations
in Born approximation with damping.

The same kind of exact cancellation is responsible for the damping of
only ``half the electrons'' in $n_{k,q}$, $n_{k,q,q'}$, and
$N_{k,q,q'}$. These correlations enter the differential equations for
correlations of lower order only in the form $\widetilde{n}_{k,q}(t)
:= n_{k,q}(t) + \sum_{q'>0} ( n_{k,q,q'}(t) + n_{k+q',q,q'}(t) ) =
\sum_{k'} \left\{ \langle \psi_{k'}^{\dag} \psi_{k-q}^{\dag} \psi_{k}
\psi_{k'-q} \rangle_t^c + \langle \psi_{k'-q}^{\dag} \psi_{k}^{\dag}
\psi_{k-q} \psi_{k'} \rangle_t^c \right\}$ and $\widetilde{N}_{k,q}(t)
:= - \sum_{q'>0} ( N_{k,q,q'}(t) - N_{k+q',q,q'}(t) ) = i \sum_{k'}
\left\{ \langle \psi_{k'}^{\dag} \psi_{k-q}^{\dag} \psi_{k}
\psi_{k'-q} \rangle_t^c - \langle \psi_{k'-q}^{\dag} \psi_{k}^{\dag}
\psi_{k-q} \psi_{k'} \rangle_t^c \right\}$. As the non-interacting
energy is the same for all these correlations, i.e.\ vanishes
exactly due to the linear energy dispersion where momentum and
energy conservation are identical for the electrons, only the
quantities $\widetilde{n}_{k,q}$ and $\widetilde{N}_{k,q}$ must be
considered. This fact was also used in the numerical calculations
since it considerably reduces the number of differential equations
to be solved. It also allows the extension of the transport
equations by ``half an order'', i.e.\ in the next order we take into
account only certain sums of the correlations over momenta such that
the number of indices does not increase. To be specific, we consider
the correlations which arise from $K_{k,q,q'}$, $k_{k,q,q'}$,
$L_{k,q,q'}$, and $l_{k,q,q'}$ by the substitution of one or both
phonon operators $B_q$ by the density operator $\rho_q = \sum_{k'}
\psi_{k'-q}^{\dag} \psi_{k'}$. These correlations as well as the
extended transport equations are given in the Appendix.

Figs.~\ref{fig:fermi_1rep}-\ref{fig:fermi_distr80} show the numerical
results for the Tomonaga-Luttinger model. As discussed in
Ref.~\cite{MWFS} the renormalization of the phonon propagation
(Eq.~(\ref{eq:phon_modes})) is important for the Born approximation to
describe the relaxation process adequately. The time evolution of the
electron occupation number at the first phonon replica shown in
Fig.~\ref{fig:fermi_1rep} demonstrates that the kinetic equations
including two-particle correlations (long dashed line) incorporate
this renormalization in the dynamics as discussed above. For short
times these equations give the same improvement for the time evolution
of the second phonon replica in comparison with the Born approximation
as in the one-electron problem (Fig.~\ref{fig:fermi_2rep}). For
larger times the agreement with the exact solution is worse because
the important renormalization of the phonon propagation is not
included in the differential equations for $K_{k,q,q'}$, $k_{k,q,q'}$,
$L_{k,q,q'}$, and $l_{k,q,q'}$. This is achieved by the extension of
the kinetic equations by the correlations in which the phonon
operators $B_q$ are substituted by the electron density operators
$\rho_q$ (Eqs.~(\ref{eq:def_pr})-(\ref{eq:def_st})). The improvement
by this dynamical renormalization can be seen in
Fig.~\ref{fig:fermi_2rep} where the results of the extended kinetic
equations (named ``highest order'', dot-dashed line) are in good
agreement with the exact solution. Figs.~\ref{fig:fermi_distr30} and
\ref{fig:fermi_distr80} show the electron distribution function in the
range of the first and second phonon replica for two different times.
Observe the excellent agreement between the results of the kinetic
equations of ``highest order'' (long dashed line) and the exact
solution (solid line) even for large times
(Fig.~\ref{fig:fermi_distr80}).

%
%
%
%
%

\subsection{Initial correlations}
\label{sec:init_corr}

\begin{figure}[htbp]
  \begin{center}
    \leavevmode
    \epsfig{file=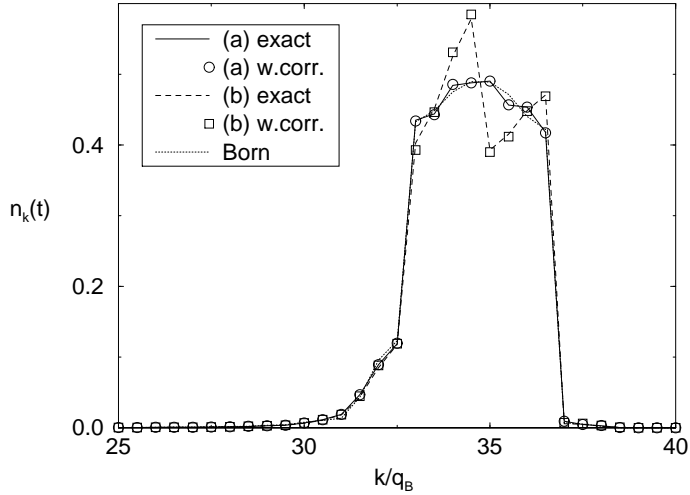,height=10cm,angle=270}
  \end{center}
  \caption{The electron distribution function at $\omega_0 t=5$ for
    a four-electron system with $\nu = 1/{2}$, $\Gamma = 0.01$,  
    and $q_c/q_B= 10$. The system was in two different initial states
    (a) and (b) described in the text with the same initial
    distribution functions. At time $t=0$ the electron distribution
    function was 0.5 at $k = 33$--$36.5 q_B$ and vanished elsewhere.
    The exact solution is compared with the results from
    the kinetic equations in Born approximation and from the improved
    equations including two-particle correlations.}
  \label{fig:init5}
\end{figure}

Finally, we discuss the role of initial correlations for the dynamics.
The initial correlations enter as initial values of the kinetic
equations. In order to study the capability of the kinetic equations
(\ref{eq:EOM_nk})-(\ref{eq:EOM_L}) to describe the effect of initial
correlations we consider different initial states $\rho_0$ inducing
identical initial distribution functions $n_k(t_0)$ and $N_q(t_0)$.
Since the dynamics is exactly soluble for $\rho_0= \vert k_1,\ldots,
k_N\rangle \langle k_1,\ldots, k_N \vert$ the initial states were
taken as convex combinations of states of this form. The exact
solution is described elsewhere \cite{WS}.  In the example shown in
Fig.~\ref{fig:init5} the two initial states considered are
\begin{eqnarray*}
  \mbox{(a)}\quad
  \rho_0 &=& \frac{1}{2} 
  \left\{ \vert k_0,k_0-q_B,k_0-2q_B,k_0-3q_B \rangle
    \langle \dots \vert
  \right.
  \\  &&
  \left. +  \vert k_0-\frac{1}{2} q_B,k_0-\frac{3}{2}
    q_B,k_0-\frac{5}{2} q_B ,k_0-\frac{7}{2} q_B \rangle
    \langle \dots  \vert
  \right\}
  \\
  \mbox{(b)}\quad
  \rho_0 &=& \frac{1}{2} 
  \left\{ \vert k_0,k_0-\frac{1}{2}q_B,k_0-q_B,k_0-\frac{3}{2}q_B \rangle
    \langle \dots \vert
  \right.
  \\ &&
  \left. +  \vert  k_0-2q_B,k_0-\frac{5}{2} q_B,k_0-3q_B
    ,k_0-\frac{7}{2} q_B \rangle 
    \langle \dots \vert 
  \right\}
\end{eqnarray*}
with a very small ``resonant'' phonon momentum $q_B= 2\cdot
\frac{2\pi}{L}$ in order to make the effect more pronounced. As the
electron and phonon distribution functions $n_k(t_0)$ and $N_q(t_0)$
as well as the phonon assisted density matrices $h_{k,q}(t_0)$ and
$H_{k,q}(t_0)$ are the same for both cases the Born approximation
yields identical results whereas the kinetic equations including
two-particle correlations provide an excellent approximation for the
exact time evolution (Fig.~\ref{fig:init5}).

%
%
%
%
%

\section{Summary}

We studied kinetic equations for electron-phonon systems beyond the
Born approximation by the inclusion of many-particle correlations. The
dynamics is described by a system of differential equations for the
distribution functions and single-time correlations. The results from
the approximate kinetic equations were compared with the exact solutions
for one-dimensional electron-phonon models. In contrast to real-time
Green's function techniques, which have definite advantages in the
context of quantum field theories, equations for retarded/advanced
Green's 
functions are not needed.

For systems with intrinsic damping, like the one-dimensional
electron-phonon models studied, already the inclusion of two-particle
correlations leads to a significant improvement in comparison with the
Born approximation. The renormalization of the electron and phonon
propagation, which is important to reduce the occurrence of unphysical
results in the Born approximation, is automatically incorporated in
the dynamics. There is no need for the Kadanoff-Baym ansatz. In
addition to the renormalization of the propagators, the improved
kinetic equations also correctly describe coherent two-phonon
processes. We studied in detail the one-electron polaron model and the
Tomonaga-Luttinger model.

Finally, the ability of the improved kinetic equations to describe the
effect of initial correlations was shown by comparison with exact
solutions for the many-electron case.
 
\acknowledgments
This work was financially supported by the Deutsche
Forschungsgemeinschaft (SFB 345 ``Festk\"{o}rper weit weg vom
Gleichgewicht''). Parts of the numerical calculations were done on
machines of the RRZN (Regionales Rechenzentrum f\"ur Niedersachsen).

%
%
%
%
%

\appendix
\section*{}

In this Appendix we present the extension of the kinetic equations
mentioned in Sec.~\ref{sec:TL}. In addition to the two-particle
correlations, we consider the following correlations 
\begin{eqnarray}
  \label{eq:def_pr}
  p_{k,q,q'}(t) &=& i \sum_{k'}
  \left( \langle B_{q'}^{\dag} \psi_{k'-q}^{\dag} \psi_{k-q'}^{\dag}
    \psi_{k-q} \psi_{k'} \rangle_t^c -
    \langle B_{q'} \psi_{k'}^{\dag} \psi_{k-q}^{\dag}
    \psi_{k-q'} \psi_{k'-q} \rangle_t^c
  \right)
  \\
  P_{k,q,q'}(t) &=& \sum_{k'}
  \left( \langle B_{q'}^{\dag} \psi_{k'-q}^{\dag} \psi_{k-q'}^{\dag}
    \psi_{k-q} \psi_{k'} \rangle_t^c +
    \langle B_{q'} \psi_{k'}^{\dag} \psi_{k-q}^{\dag}
    \psi_{k-q'} \psi_{k'-q} \rangle_t^c
  \right)
  \\
  r_{k,q,q'}(t) &=& i \sum_{k'}
  \left( \langle B_{q'}^{\dag} \psi_{k'}^{\dag} \psi_{k-q-q'}^{\dag}
    \psi_{k} \psi_{k'-q} \rangle_t^c -
    \langle B_{q'} \psi_{k'-q}^{\dag} \psi_{k}^{\dag}
    \psi_{k-q-q'} \psi_{k'} \rangle_t^c
  \right)
  \\ 
  R_{k,q,q'}(t) &=&  \sum_{k'}
  \left( \langle B_{q'}^{\dag} \psi_{k'}^{\dag} \psi_{k-q-q'}^{\dag}
    \psi_{k} \psi_{k'-q} \rangle_t^c +
    \langle B_{q'} \psi_{k'-q}^{\dag} \psi_{k}^{\dag}
    \psi_{k-q-q'} \psi_{k'} \rangle_t^c
  \right)
  \quad,
\end{eqnarray}
which arise from $K_{k,q,q'}$, $k_{k,q,q'}$, $L_{k,q,q'}$, and
$l_{k,q,q'}$ by the substitution of one phonon operator $B_q$ by the
density operator $\rho_q = \sum_{k'} \psi_{k'-q}^{\dag} \psi_{k'}$, and
\begin{eqnarray}
  s_{k,q,q'}(t) &=& \sum_{k',k''}
  \left( \langle \psi_{k''}^{\dag} \psi_{k'-q}^{\dag} \psi_{k-q'}^{\dag}
    \psi_{k-q} \psi_{k'} \psi_{k''-q'} \rangle_t^c +
    \langle \psi_{k''-q'}^{\dag} \psi_{k'}^{\dag} \psi_{k-q}^{\dag}
    \psi_{k-q'} \psi_{k'-q} \psi_{k''} \rangle_t^c
  \right)
  \\ 
  S_{k,q,q'}(t) &=& i \sum_{k',k''}
  \left( \langle \psi_{k''}^{\dag} \psi_{k'-q}^{\dag} \psi_{k-q'}^{\dag}
    \psi_{k-q} \psi_{k'} \psi_{k''-q'} \rangle_t^c -
    \langle \psi_{k''-q'}^{\dag} \psi_{k'}^{\dag} \psi_{k-q}^{\dag}
    \psi_{k-q'} \psi_{k'-q} \psi_{k''} \rangle_t^c
  \right)
  \\ 
  t_{k,q,q'}(t) &=&  \sum_{k',k''}
  \left( \langle \psi_{k''}^{\dag} \psi_{k'}^{\dag} \psi_{k-q-q'}^{\dag}
    \psi_{k} \psi_{k'-q}  \psi_{k''-q'} \rangle_t^c +
    \langle \psi_{k''-q'}^{\dag} \psi_{k'-q}^{\dag} \psi_{k}^{\dag}
    \psi_{k-q-q'} \psi_{k'}  \psi_{k''} \rangle_t^c
  \right)
  \\ 
  T_{k,q,q'}(t) &=& i \sum_{k',k''}
  \left( \langle \psi_{k''}^{\dag} \psi_{k'}^{\dag} \psi_{k-q-q'}^{\dag}
    \psi_{k} \psi_{k'-q}  \psi_{k''-q'} \rangle_t^c -
    \langle \psi_{k''-q'}^{\dag} \psi_{k'-q}^{\dag} \psi_{k}^{\dag}
    \psi_{k-q-q'} \psi_{k'}  \psi_{k''} \rangle_t^c
  \right)
  \quad,
  \label{eq:def_st}
\end{eqnarray}
in which the second phonon operator is also substituted by the
corresponding density operator.

The extension of our transport equations by these correlations reads
\begin{eqnarray}
  \label{eq:EOM_nkl}
  \frac{d}{dt} \widetilde{n}_{k,q} &=& \ldots + \sum_{q'>0} g_{q'}
  \left\{ p_{k,q,q'} - p_{k+q',q,q'} + r_{k,q,q'} - r_{k+q',q,q'}
  \right\}
  \\
  \frac{d}{dt} \widetilde{N}_{k,q} &=& \ldots + \sum_{q'>0} g_{q'}
  \left\{ P_{k,q,q'} - P_{k+q',q,q'} - R_{k,q,q'} + R_{k+q',q,q'}
  \right\}
  \\
  \left( \frac{d}{dt} +2\gamma
  \right) k_{k,q,q'} &=& \ldots - g_{q'} p_{k,q',q} - g_{q} p_{k,q,q'}
  \\
  \left( \frac{d}{dt} +2\gamma
  \right) K_{k,q,q'} &=& \ldots + g_{q'} P_{k,q',q} - g_{q} P_{k,q,q'}
  \\
  \left( \frac{d}{dt} +2\gamma
  \right) l_{k,q,q'} &=& \ldots + g_{q'} r_{k,q',q} + g_{q} r_{k,q,q'}
  \\
  \left( \frac{d}{dt} +2\gamma
  \right) L_{k,q,q'} &=& \ldots - g_{q'} R_{k,q',q} - g_{q} R_{k,q,q'}
\end{eqnarray}
where the dots stand for the terms in these equations written down so
far,
\begin{eqnarray}
  \label{eq:EOM_pr}
  \left( \frac{d}{dt} +2\gamma
  \right) p_{k,q,q'} &=& -(\omega_{q'}-v_F q') P_{k,q,q'}
  \\ \nonumber{} &&
  + g_{q'}
  \left\{ -
    \left(1+N_{q'}-n_{k-q'}
    \right) \widetilde{n}_{k,q} 
    +
    \left( N_{q'}+n_{k-q}
    \right) \widetilde{n}_{k-q',q}
  \right\}
  \\ \nonumber{} &&
  + g_{q} \frac{1}{2} 
  \left\{ (h_{k-q,q'}-h_{k,q'}) \sum_{k'} h_{k',q} -
    (H_{k-q,q'}-H_{k,q'}) \sum_{k'} H_{k',q}
  \right\}
  \\ \nonumber{} &&
  + g_{q}
  \sum_{k'}(n_{k'-q}-n_{k'}) k_{k,q,q'}
  - g_{q'} s_{k,q,q'}
  \\
   \left( \frac{d}{dt} +2\gamma
  \right) P_{k,q,q'} &=& (\omega_{q'}-v_F q') p_{k,q,q'}
  \\ \nonumber{} &&
  + g_{q'}
  \left\{ -
    \left(1+N_{q'}-n_{k-q'}
    \right) \widetilde{N}_{k,q} 
    +
    \left( N_{q'}+n_{k-q}
    \right) \widetilde{N}_{k-q',q}
  \right\}
  \\ \nonumber{} &&
  + g_{q} \frac{1}{2} 
  \left\{ (h_{k-q,q'}-h_{k,q'}) \sum_{k'} H_{k',q} +
    (H_{k-q,q'}-H_{k,q'}) \sum_{k'} h_{k',q}
  \right\}
  \\ \nonumber{} &&
  + g_{q}
  \sum_{k'}(n_{k'-q}-n_{k'}) K_{k,q,q'}
 + g_{q'} S_{k,q,q'}
  \\
  \left( \frac{d}{dt} +2\gamma
  \right) r_{k,q,q'} &=& -(\omega_{q'}-v_F q') R_{k,q,q'}
  \\ \nonumber{} &&
  + g_{q'}
  \left\{ -
    \left( 1+N_{q'}-n_{k-q-q'}
    \right) \widetilde{n}_{k,q} 
    +
    \left( N_{q'}+n_{k}
    \right) \widetilde{n}_{k-q',q}
  \right\}
  \\ \nonumber{} &&
  + g_{q} \frac{1}{2} 
  \left\{ (h_{k-q,q'}-h_{k,q'}) \sum_{k'} h_{k',q} +
    (H_{k-q,q'}-H_{k,q'}) \sum_{k'} H_{k',q}
  \right\}
  \\ \nonumber{} &&
  - g_{q}
  \sum_{k'}(n_{k'-q}-n_{k'}) l_{k,q,q'}
   - g_{q'} t_{k,q,q'}
  \\
  \left( \frac{d}{dt} +2\gamma
  \right) R_{k,q,q'} &=& (\omega_{q'}-v_F q') r_{k,q,q'}
  \\ \nonumber{} &&
  + g_{q'}
  \left\{ 
    \left( 1+N_{q'}-n_{k-q-q'}
    \right) \widetilde{N}_{k,q} 
    -
    \left( N_{q'}+n_{k}
    \right) \widetilde{N}_{k-q',q}
  \right\}
  \\ \nonumber{} &&
  + g_{q} \frac{1}{2} 
  \left\{ -(h_{k-q,q'}-h_{k,q'}) \sum_{k'} H_{k',q} +
    (H_{k-q,q'}-H_{k,q'}) \sum_{k'} h_{k',q}
  \right\}
  \\ \nonumber{} &&
  + g_{q}
  \sum_{k'}(n_{k'-q}-n_{k'}) L_{k,q,q'}
  + g_{q'} T_{k,q,q'}
  \\
\end{eqnarray}
and last but not least
\begin{eqnarray}
  \label{eq:EOM_st}
  \left( \frac{d}{dt} +2\gamma
  \right) s_{k,q,q'} &=& g_q
  \left\{ \sum_{k'} (n_{k'-q}-n_{k'}) p_{k,q',q}
  \right.
  \\ \nonumber{} &+&
  \left.  \frac{1}{2}
    \left[ (\widetilde{N}_{k-q,q'} -\widetilde{N}_{k,q'}) \sum_{k'}
      H_{k',q} + (\widetilde{n}_{k-q,q'} -\widetilde{n}_{k,q'}) \sum_{k'}
      h_{k',q} 
    \right]
  \right\} + (q \leftrightarrow q')
  \\
  \left( \frac{d}{dt} +2\gamma
  \right) S_{k,q,q'} &=& g_q
  \left\{ \sum_{k'} (n_{k'-q}-n_{k'}) P_{k,q',q}
  \right.
  \\ \nonumber{} &+&
  \left.  \frac{1}{2}
    \left[ (\widetilde{N}_{k-q,q'} -\widetilde{N}_{k,q'}) \sum_{k'}
      h_{k',q} - (\widetilde{n}_{k-q,q'} -\widetilde{n}_{k,q'}) \sum_{k'}
      H_{k',q} 
    \right]
  \right\} - (q \leftrightarrow q')
  \\
  \left( \frac{d}{dt} +2\gamma
  \right) t_{k,q,q'} &=& g_q
  \left\{ \sum_{k'} (n_{k'-q}-n_{k'}) r_{k,q',q}
  \right.
  \\ \nonumber{} &+&
  \left.  \frac{1}{2}
    \left[- (\widetilde{N}_{k-q,q'} -\widetilde{N}_{k,q'}) \sum_{k'}
      H_{k',q} + (\widetilde{n}_{k-q,q'} -\widetilde{n}_{k,q'}) \sum_{k'}
      h_{k',q} 
    \right]
  \right\} + (q \leftrightarrow q')
  \\
  \left( \frac{d}{dt} +2\gamma
  \right) T_{k,q,q'} &=& g_q
  \left\{- \sum_{k'} (n_{k'-q}-n_{k'}) R_{k,q',q}
  \right.
  \\ \nonumber{} &+&
  \left.  \frac{1}{2}
    \left[ (\widetilde{N}_{k-q,q'} -\widetilde{N}_{k,q'}) \sum_{k'}
      h_{k',q} + (\widetilde{n}_{k-q,q'} -\widetilde{n}_{k,q'}) \sum_{k'}
      H_{k',q} 
    \right]
  \right\} + (q \leftrightarrow q')
  \quad.
\end{eqnarray}

The same discussion as for the damping of the phonon assisted density
matrices (Eqs.~(\ref{eq:damp_start})-(\ref{eq:Dyson_sol})) shows that
the terms $p_{\dots}$, $P_{\dots}$, $r_{\dots}$, and $R_{\dots}$ in
the time evolution of $\widetilde{n}_{k,q}$ and $\widetilde{N}_{k,q}$
lead to a damping of $\widetilde{n}_{k,q}$ and $\widetilde{N}_{k,q}$
by the same factor $2\gamma$, as mentioned in Sec.~\ref{sec:pol}. In
general, in all correlations containing a density operator $\rho_q$,
i.e.\ $\sum_{k'} \langle \psi^{\dag}_{k'-q} \psi_{k'}
\ldots\rangle^c_t$, these electron propagations are not damped.

%
%
%
%
%

%
%
%
%
%

\begin{figure}[htbp]
  \begin{center}
    \leavevmode
    \epsfig{file=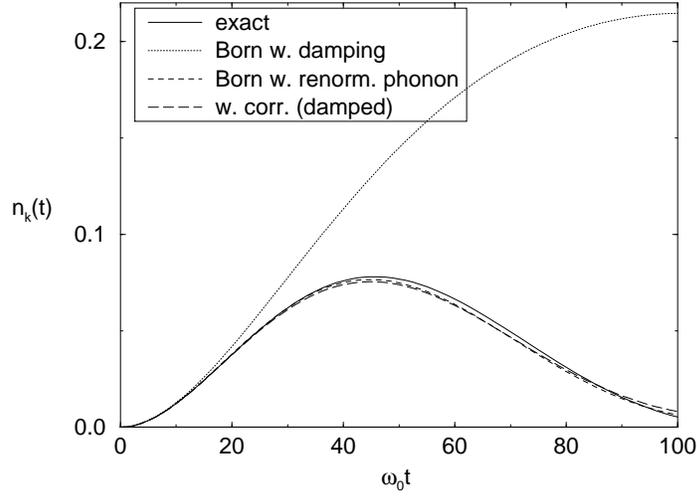,height=10cm,angle=270}
  \end{center}
  \caption{Time evolution of the electron occupation at the first
    phonon replica ($k=k_0-q_B$) for the Tomonaga-Luttinger model with
    $\nu = 1/{20}$, $\Gamma = 0.003$, and $q_c/q_B= 3$.
    The excited electron was initially at $k_0= 6 q_B$
    ($k_F=0$). The exact solution is compared with the results from
    the kinetic equations in Born approximation with and without
    renormalization of the phonon propagation and the improved
    equations including two-particle correlations.}
  \label{fig:fermi_1rep}
\end{figure}

\begin{figure}[htbp]
  \begin{center}
    \leavevmode
    \epsfig{file=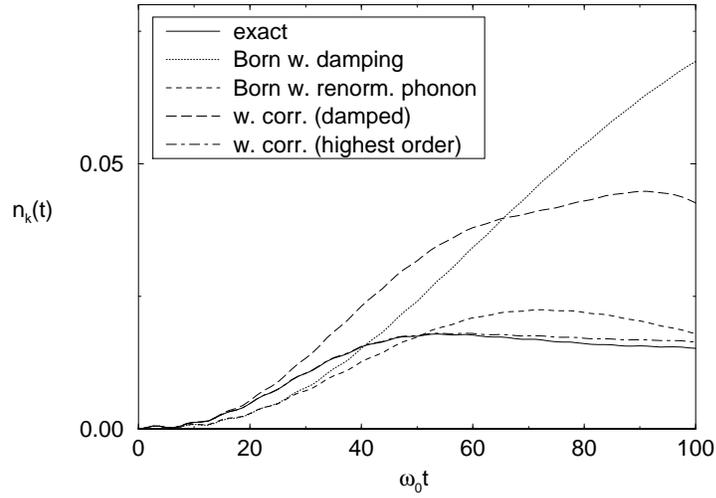,height=10cm,angle=270}
  \end{center}
  \caption{Time evolution of the electron occupation at the second
    phonon replica ($k=k_0-2 q_B$) for the system of
    Fig.~\ref{fig:fermi_1rep}. In addition to Fig.~\ref{fig:fermi_1rep}
    the kinetic equations are extended by ``half an order''.
    }
  \label{fig:fermi_2rep}
\end{figure}

\begin{figure}[htbp]
  \begin{center}
    \leavevmode
    \epsfig{file=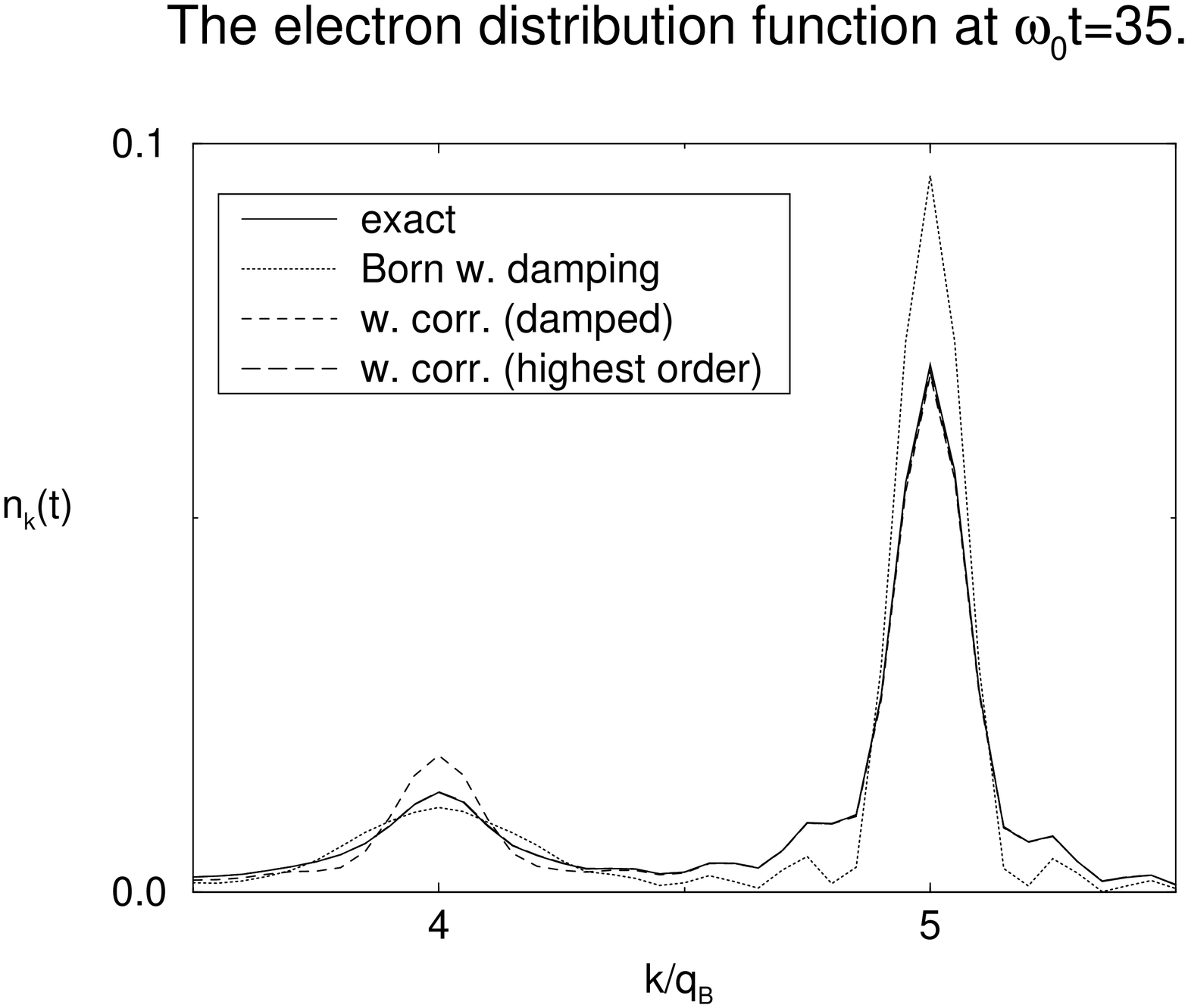,height=10cm,angle=270}
  \end{center}
  \caption{The electron distribution function at $\omega_0 t=35$ for
    the system of Fig.~\ref{fig:fermi_1rep}. Shown are the first and
    second phonon replica. The electron was initially at $k_0=
    6 q_B$.} 
  \label{fig:fermi_distr30}
\end{figure}

\begin{figure}[htbp]
  \begin{center}
    \leavevmode
    \epsfig{file=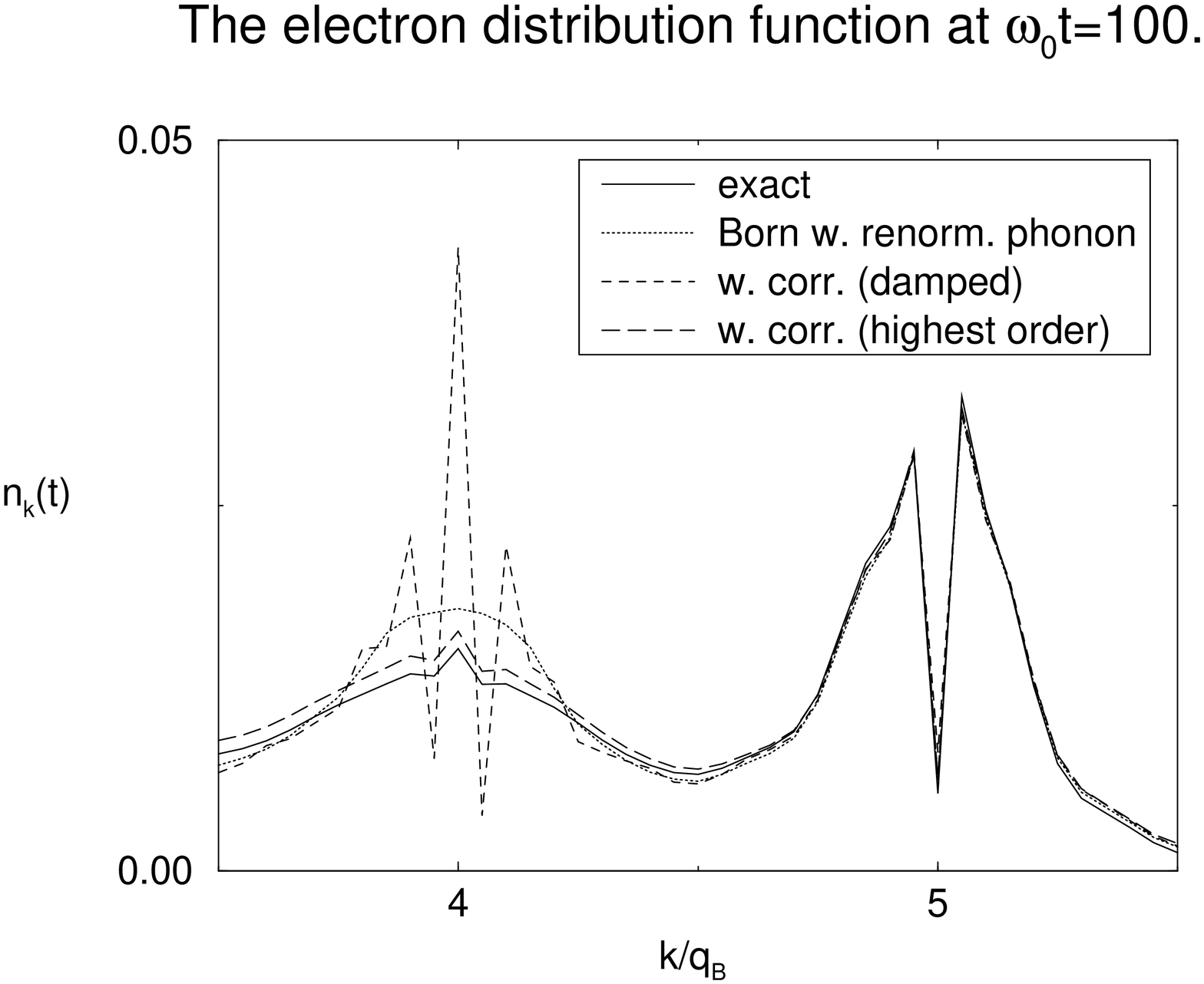,height=10cm,angle=270}
  \end{center}
  \caption{The same as in Fig.~\ref{fig:fermi_distr30}, but for the time
    $\omega_0 t=100$. The Born approximation without renormalization of
    the phonon propagation is omitted as it does not yield acceptable
    results at larger times (see Fig.~\ref{fig:fermi_1rep}).}
  \label{fig:fermi_distr80}
\end{figure}

\end{document}